

\def\q{q{\bar q}}

\def\lsim{\raise0.3ex\hbox{$<$\kern-0.75em\raise-1.1ex\hbox{$\sim$}}}
\def\gsim{\raise0.3ex\hbox{$>$\kern-0.75em\raise-1.1ex\hbox{$\sim$}}}

\newcount\REFERENCENUMBER\REFERENCENUMBER=0
\def\REF#1{\expandafter\ifx\csname RF#1\endcsname\relax
               \global\advance\REFERENCENUMBER by 1
               \expandafter\xdef\csname RF#1\endcsname
                   {\the\REFERENCENUMBER}\fi}
\def\reftag#1{\expandafter\ifx\csname RF#1\endcsname\relax
               \global\advance\REFERENCENUMBER by 1
               \expandafter\xdef\csname RF#1\endcsname
                      {\the\REFERENCENUMBER}\fi
             \csname RF#1\endcsname\relax}
\def\ref#1{\expandafter\ifx\csname RF#1\endcsname\relax
               \global\advance\REFERENCENUMBER by 1
               \expandafter\xdef\csname RF#1\endcsname
                      {\the\REFERENCENUMBER}\fi
             [\csname RF#1\endcsname]\relax}
\def\refto#1#2{\expandafter\ifx\csname RF#1\endcsname\relax
               \global\advance\REFERENCENUMBER by 1
               \expandafter\xdef\csname RF#1\endcsname
                      {\the\REFERENCENUMBER}\fi
           \expandafter\ifx\csname RF#2\endcsname\relax
               \global\advance\REFERENCENUMBER by 1
               \expandafter\xdef\csname RF#2\endcsname
                      {\the\REFERENCENUMBER}\fi
             [\csname RF#1\endcsname--\csname RF#2\endcsname]\relax}
\def\refs#1#2{\expandafter\ifx\csname RF#1\endcsname\relax
               \global\advance\REFERENCENUMBER by 1
               \expandafter\xdef\csname RF#1\endcsname
                      {\the\REFERENCENUMBER}\fi
           \expandafter\ifx\csname RF#2\endcsname\relax
               \global\advance\REFERENCENUMBER by 1
               \expandafter\xdef\csname RF#2\endcsname
                      {\the\REFERENCENUMBER}\fi
            [\csname RF#1\endcsname,\csname RF#2\endcsname]\relax}
\def\refss#1#2#3{\expandafter\ifx\csname RF#1\endcsname\relax
               \global\advance\REFERENCENUMBER by 1
               \expandafter\xdef\csname RF#1\endcsname
                      {\the\REFERENCENUMBER}\fi
           \expandafter\ifx\csname RF#2\endcsname\relax
               \global\advance\REFERENCENUMBER by 1
               \expandafter\xdef\csname RF#2\endcsname
                      {\the\REFERENCENUMBER}\fi
           \expandafter\ifx\csname RF#3\endcsname\relax
               \global\advance\REFERENCENUMBER by 1
               \expandafter\xdef\csname RF#3\endcsname
                      {\the\REFERENCENUMBER}\fi
[\csname RF#1\endcsname,\csname RF#2\endcsname,\csname
RF#3\endcsname]\relax}
\def\refand#1#2{\expandafter\ifx\csname RF#1\endcsname\relax
               \global\advance\REFERENCENUMBER by 1
               \expandafter\xdef\csname RF#1\endcsname
                      {\the\REFERENCENUMBER}\fi
           \expandafter\ifx\csname RF#2\endcsname\relax
               \global\advance\REFERENCENUMBER by 1
               \expandafter\xdef\csname RF#2\endcsname
                      {\the\REFERENCENUMBER}\fi
            [\csname RF#1\endcsname,\csname RF#2\endcsname]\relax}
\def\Ref#1{\expandafter\ifx\csname RF#1\endcsname\relax
               \global\advance\REFERENCENUMBER by 1
               \expandafter\xdef\csname RF#1\endcsname
                      {\the\REFERENCENUMBER}\fi
             [\csname RF#1\endcsname]\relax}
\def\Refto#1#2{\expandafter\ifx\csname RF#1\endcsname\relax
               \global\advance\REFERENCENUMBER by 1
               \expandafter\xdef\csname RF#1\endcsname
                      {\the\REFERENCENUMBER}\fi
           \expandafter\ifx\csname RF#2\endcsname\relax
               \global\advance\REFERENCENUMBER by 1
               \expandafter\xdef\csname RF#2\endcsname
                      {\the\REFERENCENUMBER}\fi
            [\csname RF#1\endcsname--\csname RF#2\endcsname]\relax}
\def\Refand#1#2{\expandafter\ifx\csname RF#1\endcsname\relax
               \global\advance\REFERENCENUMBER by 1
               \expandafter\xdef\csname RF#1\endcsname
                      {\the\REFERENCENUMBER}\fi
           \expandafter\ifx\csname RF#2\endcsname\relax
               \global\advance\REFERENCENUMBER by 1
               \expandafter\xdef\csname RF#2\endcsname
                      {\the\REFERENCENUMBER}\fi
        [\csname RF#1\endcsname,\csname RF#2\endcsname]\relax}
\def\refadd#1{\expandafter\ifx\csname RF#1\endcsname\relax
               \global\advance\REFERENCENUMBER by 1
               \expandafter\xdef\csname RF#1\endcsname
                      {\the\REFERENCENUMBER}\fi \relax}

%

\def\NP{{ Nucl.\ Phys.\ }}
\def\PL{{ Phys.\ Lett.\ }}
\def\PR{{ Phys.\ Rev.\ }}

\def\PRL{{ Phys.\ Rev.\ Lett.\ }}

\def\ZP{{ Z.\ Phys.\ }}

%

\magnification=1200
\hsize=16.0truecm
\vsize=24.5truecm
\baselineskip=13pt
\def\q{\q{\bar q}}
\def\Le{\it l^+ l^-}
\def\g{\gamma^*}
\def\h{h^+ h^-}
\pageno=0
\input macro-hs
{}~~~
\hfill CERN-TH.7399/94\par
\hfill BI-TP 94/39\par
\vskip 2.5truecm
\centerline{\bf CHIRAL DYNAMICS, DECONFINEMENT}
\medskip
\centerline{\bf AND THERMAL DILEPTONS}
\vskip 1.5 truecm
\centerline{D.\ Kharzeev and H.\ Satz}
\bigskip\medskip
\centerline{Theory Division, CERN, CH-1211 Geneva, Switzerland}
\centerline{and}
\centerline{Fakult\"at f\"ur Physik, Universit\"at Bielefeld,
D-33501 Bielefeld, Germany}
\vskip 2 truecm
\centerline{\bf Abstract:}
We show that in a hot hadron gas, chiral symmetry restoration and
deconfinement lead to a strong suppression of dilepton
production by pion annihilation in the continuum above the
$\rho-\omega-\phi$ peak.
The suppression due to deconfinement
persists also for annihilation reactions involving heavier mesons.
\par\vfill\noindent
CERN-TH.7399/94\par\noindent
BI-TP 94/39\par\noindent
August 1994
\eject

Thermal dileptons were among the first probes to be suggested for the
study of strongly interacting matter \ref{Feinberg}. Their small
electromagnetic cross section allows them to escape from the medium
essentially unmodified and, in contrast to hadrons, they can therefore
provide information even about the early evolution stages of this
medium. Thermal dilepton spectra decrease exponentially for large mass
$M$, ${\rm exp}(-M/T)$, and if measurable, they would therefore serve
as ideal thermometers of early matter \refs{Kajantie}{Gyulassy}. To use
them also as a probe for the deconfinement or
chiral symmetry aspects of such matter,
we have to consider in more detail how they can be
produced \ref{Domokos}, since the exponential thermal form will hold for
dileptons from hadronic ($\pi^+ \pi^-\to \g \to \Le$) as well as from
partonic ($\q \to \g \to \Le$) interactions. We shall begin with some
comments on the difference between these two types of annihilation.
\par
A timelike virtual photon $\g$ of mass $\sqrt{q^2}$=2 GeV
is characterized by a spatial scale of about 0.1 fm.
For two hadrons of the typical size of 1 fm to annihilate into a
ten times smaller entity is evidently rather unlikely; a good
illustration of this is the minute cross section for $p {\bar p} \to \Le$
\ref{data}. In parton language, such a process requires the
simultaneous
annihilation of $n$ valence quark-antiquark pairs ($n$=2 for mesons,
$n$=3 for
baryons), and by quark counting rules \ref{y}, this is suppressed by a
factor
$$
F_h \sim {1\over (q^2)^{n-1}}. \eqno(1)
$$
This factor is a direct consequence of confinement;
the power behaviour (1) arises from the propagators of the hard
gluons needed to transfer the information about the
annihilation from one of the quark pairs to the
$(n-1)$ others.
The relevant diagram for pion annihilation is shown in Fig.\ 1.
\vskip 7.5truecm
\centerline{Figure 1: Pion annihilation in QCD}
\bigskip
In a deconfined medium, different parton interactions are not
correlated, so that the annihilation is now described by the elementary
electromagnetic process $\q \to \g \to \Le$,
whose cross section is given by
$$
\sigma_{\q}(M) \sim \left( {\alpha^2 \over M^2} \right). \eqno(2)
$$
In contrast, we have
$$
\sigma_{\h}(M) \sim F^2_h(M) \left( {\alpha^2 \over M^2} \right) \eqno(3)
$$
for hadron-antihadron annihilation. Because of their small size, high
mass thermal dileptons thus quite generally reflect the parton structure
of the medium from which they originate; it is the presence or absence
of the suppression factor $F_h(M)$ which indicates if this medium is in
a confined or deconfined state. Let us consider $F_h$ in more detail
for the specific case $\pi^+ \pi^-\to \g \to \Le$.
\par
The vector meson dominance model (VDM) leads to the $\rho$-pole form
$$
F_{\pi}^2 \simeq {m_{\rho}^4 \over (q^2 - m_{\rho}^2)^2 + m_{\rho}^2
\Gamma_{\rho}^2 }, \eqno(4)
$$
where $\Gamma_{\rho}\simeq$ 155 MeV is
the $\rho \to \pi^+ \pi^-$ decay width.
For large $q^2$, this form yields $F_{\pi}^2 \simeq m_{\rho}^4/(q^2)^2$
and is thus in fact in accord with the counting rule result (1).
Although widely used for the hadronic dilepton spectrum also at high
masses, its proportionality to the $\rho$-mass is difficult to justify
far away from the $\rho$-pole.
Even the agreement of its functional form at large $q^2$ with that from
the counting rules appears to be fortuitous, as can be seen
by considering the (squared) proton form factor. In
the naive VDM it also decreases as $(1/q^2)^2$, while
the counting rules lead to a much steeper $(1/q^2)^4$ fall-off,
which is supported by data from $p{\bar p}$ annihilation into dileptons
\ref{data} for positive $q^2$ and from proton form factor measurements
in $e-p$ scattering \ref{data-} for negative $q^2$.
\par
In QCD, the pion form factor can
be calculated for large $q^2$ \refss{CZ1}{CZ2}{Braun}; it is given by
$$
F_{\pi}(q^2) \simeq {32 \pi  \over 9} {\alpha_s(q^2) \over {q^2}}
 |f_{\pi}|^2  \left| \int_{-1}^1 {d\xi \over {1 - \xi^2}}
\ \varphi_{\pi}(\xi) \right|^2, \eqno(5)
$$
where $f_{\pi}\simeq 133$ MeV is the pion decay constant determined
by the $\pi \to \mu \nu$ decay. It reflects the behaviour of the pion
wave function at the scale of weak interactions ($\sim M_W^{-1}$).
The function $\varphi_{\pi}(\xi)$ describes the longitudinal momentum
distribution of the valence quarks in the pion; it is normalized to
unity,
$$
\int_{-1}^1 d\xi \  \varphi_{\pi}(\xi) = 1. \eqno(6)
$$
Here $\xi = x_1 - x_2$, with $x_1$ and $x_2$ denoting
the fractions of the pion momentum carried by quark and
antiquark. Eq.\ (5) separates nicely into a ``hard" part $\alpha_s/q^2$
from the highly virtual gluon exchange, and a remaining ``soft" part
determined by the non-perturbative properties of the pion wave function.
The distribution $\varphi_{\pi}(\xi)$ can be calculated in
perturbative QCD for $q^2 \to \infty$ \ref{CZ1} and is given
by
$$
\varphi_{\pi}(\xi) = {3\over 4}(1-\xi^2). \eqno(7)
$$
It has a maximum at $\xi=0$, where each of the valence quarks carries
half of the pion momentum, with a dispersion $\langle \xi^2 \rangle =
0.2$.
In the $q^2$ region\footnote{*}{
For finite $q^2$ and $\xi \to 1$, the validity of the underlying
factorisation assumption is still not completely clear \ref{Llew}.}
of most interest to us, $q^2 \simeq$ 2 - 10 GeV$^2$,
non-perturbative effects are generally expected to make
the distribution broader \refs{CZ2}{Braun}. A solution of moment sum
rules in QCD \ref{CZ2} gives
$$
\varphi_{\pi}(\xi) = {15\over 4}(1-\xi^2) ~\xi^2, \eqno(8)
$$
which implies $\langle \xi^2 \rangle \simeq 0.4$. On the
other hand, in the limit of no binding, when the momentum of the pion
becomes simply the sum of the two valence quark momenta,
$\varphi_{\pi}(\xi)$ is a $\delta$-function around $\xi=0$,
$$
\varphi_{\pi}(\xi) = \delta(\xi), \eqno(9)
$$
since now each quark carries half of the momentum.
\par
After this brief discussion of the pion form factor at large $q^2$,
let us now see how pion annihilation into a dilepton pair can be
modified in a hot medium. With increasing temperature, we expect
deconfinement (at $T=T_c$) and chiral symmetry restoration (at
$T=T_{\chi}$). Lattice calculations so far give $T_c=T_{\chi}$
\ref{T_c}, but they can be carried out only for vanishing baryon
number density. It is clear that even below these temperatures,
the medium will already start to be modified. Such modifications
can affect the dilepton emission from a pion gas only through
the pion form factor $F_{\pi}$. Its hard part, $\alpha_s/q^2$,
does not change for meaningful temperature variations, if we
neglect a possible slow evolution of $\alpha_s$ with temperature.
The ``soft" part, however, which is determined by the value of the
pion decay constant $f_{\pi}$ and by the pion distribution function
$\varphi(\xi)$, is likely to be modified in hot hadronic matter.
\par
The pion decay constant $f_{\pi}$ is related to the chiral condensate
$\langle \bar{q}q \rangle$ through the PCAC relation
$$
f_{\pi}^2 = {- m_q \over m_{\pi}^2}~\langle {\bar q}q \rangle.
\eqno(10)
$$
The temperature dependence of $f_{\pi}$ is thus in principle affected
by both the chiral and the confinement features of the medium. In the
quark
model, $f_{\pi}$ is proportional to the $\bar{q}q$ wave function at
the origin, and this decreases when the pion becomes more losely
bound near $T_c$. Near $T_{\chi}$, the chiral condensate
$\langle \bar{q}q \rangle$ decreases, to vanish at $T_{\chi}$ in the
case of massless quarks. So far,
lattice calculations show a slow increase of $m_{\pi}$ with
temperature, together with a rather rapid drop of
$\langle \bar{q}q \rangle$ near $T_{\chi}$ \ref{Lattice}.
This would make $f_{\pi}$ decrease more rapidly than
$\langle \bar{q}q \rangle ^{1/2}$ as $T \to T_{\chi}$. The general
observation that
$f_{\pi} \to 0$ as $T \to T_{\chi}$ is in accord with the predictions of
the effective Lagrangian approach \ref{eff}, though particular
models (see \ref{Brown} and references therein) predict somewhat slower
fall-off: $f_{\pi}(T) \sim \langle \bar{q}q(T) \rangle^{1/3}$ .
In either case, however, it is clear that $f_{\pi}$ decreases
with temperature. For the sake of argument, we shall assume the pion
mass to remain approximately temperature independent, to get the
scaling relation
$$
{f_{\pi}(T) \over f_{\pi}(0)} = \left[ {\langle q{\bar q}(T) \rangle
\over \langle q{\bar q}(0) \rangle} \right]^{1/2} \eqno(11)
$$
between pion form factor and chiral condensate.
\par
In addition, the temperature dependence of the pion form factor
will be affected by a temperature variation of the
pion distribution function $\varphi_{\pi}(\xi)$. Near the deconfinement
point, colour screening will loosen the quark-antiquark binding in the
pion. As a result, the pion wave function becomes softer
in momentum space, which leads to a narrower distribution
function $\varphi_{\pi}(\xi)$, as already noted above.
This decreases the value of the integral
$$
I(\varphi_{\pi}) = \int_{-1}^1 {d\xi \over {1 - \xi^2}} \
\varphi_{\pi}(\xi), \eqno(12)
$$
in the pion form factor (4); while the bound state form (7) leads to
$I=5/2$, the ``deconfined" $\delta$-function gives $I=1$. In addition
to the effect of chiral dynamics, deconfinement thus causes
a further decrease of dilepton production by pion annihilation.
\par
We will now estimate the temperature dependence of the pion annihilation
cross section, which by eq.\ (4) is proportional to $f_{\pi}^4$.
Assuming the validity of the chiral scaling law (11), we get
$$
\sigma_{\pi^+\pi^-}(M^2;T) \simeq \sigma_{\pi^+\pi^-}(M^2;0)\left[
{\langle \bar{q}q(T) \rangle \over \langle \bar{q}q(0) \rangle}
\right]^2 \left[{ I(T)\over I(0)}\right]^4. \eqno(13)
$$
For QCD with two flavours of massless quarks, the behaviour of the
chiral condensate near $T_{\chi}$ is conjectured to be the same as that
for the three-dimensional O(4) sigma model \ref{Wilczek}; recent lattice
studies are in accord with this conjecture \ref{Karschlat}. This
leads to
$$
{\langle \bar{q}q(T) \rangle \over \langle \bar{q}q(0) \rangle}
= \left[ 1 - {T\over T_{\chi}} \right]^{0.38} \eqno(14)
$$
and hence to a cross section decreasing as $(T-T_{\chi})^{0.76}$.
In addition, the ratio of the integrals (12) in eq.\ (13) causes a drop
by a factor $(2/5)^4 \simeq 0.03$ between the bound
state form and deconfinement.
\par
Since the masses of the (non-Goldstone) mesons may well show a
significant variation with temperature,
one may try to mimic such a decrease of the pion form factor with
increasing temperature by retaining $\rho$-pole dominance and
introducing a $\rho$-mass which decreases with
$T$.
While this may be reasonable for $q^2 \sim m_{\rho}^2$
\ref{Karsch}, it does not solve the conceptual difficulties at
large $q^2$, where the quark structure determines the form factor.
\par
Finally we comment briefly on the annihilation involving heavier
hadrons. We have seen that pion
annihilation decreases rapidly as the temperature of the medium
apporaches the point of chiral symmetry restoration and/or
deconfinement. Dilepton production
by other hadronic annihilation processes, such as $\rho~\rho \to \Le$ or
$\pi~a_1 \to \Le$, involve different form factors, which in principle
have to be analysed independently. In particular, the PCAC relation
between pion decay constant and chiral condensate is not given for the
heavier resonances. However, the
temperature dependence of the quark momentum distributions holds for all
mesons, and so the deconfinement decrease by a factor of about $10^{-2}$
remains generally applicable. --
Moreover, if it is assumed that the
form factors involving heavier mesons are proportional
\refss{Kapusta}
{Gale}{Shuryak} or closely related \ref{Ryazuddin} to the pion form
factor, then our results are immediately applicable and suppress
dilepton production for such processes as strongly as for pions.
\par
In conclusion:\ we have shown that in a hot hadron gas
near deconfinement and chiral symmetry restoration (or in the
mixed phase in case of a first order phase transition), dilepton
production by hadron annihilation is strongly suppressed.
Hence such
reactions cannot be responsible for any dilepton production
enhancement in the continuum around 1.5 Gev and higher.
\bigskip
\centerline{\bf Acknowledgements}
\medskip
We thank V. L. Chernyak for helpful remarks. The financial support of
the German Ministry for Science and Technology (BMFT) is gratefully
acknowledged.
\vfill\eject
\centerline{\bf References}
\medskip
\item{\reftag{Feinberg})}{E.L. Feinberg, Nuovo Cim. A 34 (1976) 391.}
\item{\reftag{Kajantie})}{K. Kajantie and H. I. Miettinen, \ZP C 9 (1981) 341.}
\item{\reftag{Gyulassy})}{M. Gyulassy, \NP A418 (1984) 59c.}
\item{\reftag{Domokos})}{G. Domokos and J.I. Goldman, Phys. Rev. D 23
(1981) 203.}
\item{\reftag{data})} {G. Bardin et al., Phys. Lett. B 255 (1991) 149;
B 257 (1991) 514;\hfill\break
T.A. Armstrong et al., Phys.Rev.Lett. 70 (1993) 1212.}
\item{\reftag{y})}  V.A. Matveev, R.M. Muradyan, A.N. Tavkhelidze,
Nuovo Cim. Lett. 7 (1973) 719;\hfill\break
S.J. Brodsky and G.R. Farrar, Phys.Rev.Lett. 31 (1973) 1153.
\item{\reftag{data-})} P. Stoler, Phys. Rep. 226 (1993) 103
\item{\reftag{CZ1})}{V.L. Chernyak and A.R. Zhitnitsky, JETP Lett. 25
(1977) 510; Yad. Fiz. 31 (1980) 1053;\hfill\break
G. P. Lepage and S. J. Brodsky, Phys. Lett. B 87 (1979) 359: Phys. Rev.
D 22 (1980) 2157;\hfill\break
A. V. Efremov and A. V. Radyushkin, Phys. Lett. B 94 (1980) 254.}
\item{\reftag{CZ2})} V.L. Chernyak and A.R. Zhitnitsky, Phys.Rep. 112
(1984) 173.
\item{\reftag{Braun})}{V. Braun and I. Halperin, \PL B 328 (1994) 457.}
\item{\reftag{Llew})}{N. Isgur and C. H. Llewellyn Smith, Phys. Rev.
Lett. 52 (1984) 1080; \NP B 317 (1989) 526.}
\item{\reftag{T_c})} See e.g., R. V. Gavai et al., \PL B 241 (1990)
567.
\item{\reftag{Lattice})} E. Laermann et al., \NP B Proc. Suppl. 34
(1994) 292;\hfill\break
G. Boyd et al., ``Hadron properties just before deconfinement in
quenched lattice QCD", Bielefeld Preprint BI-TP 94/42.
\item{\reftag{eff})} B.A. Campbell, J. Ellis and K.A. Olive, Phys. Lett. B 235
(1990) 325; Nucl. Phys. B 345 (1990) 57.
\item{\reftag{Brown})} C. Adami and G.E. Brown, Phys.Rep. 234 (1994) 1.
\item{\reftag{Wilczek})} R. Pisarski and F. Wilczek, \PR D 29 (1984)
338; \hfill\break
K. Rajagopal and F. Wilczek, \NP B 399 (1993) 395.
\item{\reftag{Karschlat})} F. Karsch, \PR D 49 (1994) 3791;\hfill\break
F. Karsch and E. Laermann, ``Susceptibilities, the specific heat and
a cumulant in two flavour QCD", Bielefeld Preprint BI-TP 94/29.
\item{\reftag{Karsch})}{F. Karsch, K. Redlich and L. Turko, \ZP C 60
(1993) 519.}
\item{\reftag{Kapusta})} C. Gale and J.I. Kapusta,
Nucl.Phys. B 357 (1991) 65.
\item{\reftag{Gale})}{Chungsik Song, Che Ming Ko and C. Gale, ``Role of
the $a_1$ Meson in Dilepton Production from Hot Hadronic Matter",
Preprint 1993.}
\item{\reftag{Shuryak})}{E. V. Shuryak and L. Xiong, ``Where do the
Excess Photons and Dileptons in SPS Nuclear Collisions Come from?"
Stony Brook Preprint SUNY-NTG-94-14, 1994.}
\item{\reftag{Ryazuddin})}{K. Kawarabayashi and M. Suzuki, \PRL
16 (1966) 255;\hfill\break
Riazuddin and Fayazuddin, \PR 147 (1966) 1071.
\vfill\eject\bye